\documentclass[showpacs,twocolumn,pra]{revtex4}

\usepackage{graphicx}
\usepackage{latexsym}
\usepackage{times}
\usepackage{psfrag,amsfonts}

\newcommand{\ket}[1]{|#1 \rangle}

\newcommand{\ketbra}[2]{|#1\rangle\langle#2|}
\newcommand{\average}[1]{ \langle #1  \rangle}

\newcommand{\underRarr}[1]{\begin{array}{c} #1 \\ \longrightarrow\\ \\ \end{array}}

\newcommand{\be}{\begin{equation}}
\newcommand{\ee}{\end{equation}}
\newcommand{\beq}{\begin{equation}}
\newcommand{\eeq}{\end{equation}}
\newcommand{\bae}{\begin{eqnarray}}
\newcommand{\eae}{\end{eqnarray}}

\def\CC{{\rm\kern.24em \vrule width.04em height1.46ex depth-.07ex
    \kern-.30em C}}
\def\P{{\rm I\kern-.25em P}}

\def\bbbc{{\mathchoice {\setbox0=\hbox{$\displaystyle\rm C$}\hbox{\hbox
to0pt{\kern0.4\wd0\vrule height0.9\ht0\hss}\box0}}
{\setbox0=\hbox{$\textstyle\rm C$}\hbox{\hbox
to0pt{\kern0.4\wd0\vrule height0.9\ht0\hss}\box0}}
{\setbox0=\hbox{$\scriptstyle\rm C$}\hbox{\hbox
to0pt{\kern0.4\wd0\vrule height0.9\ht0\hss}\box0}}
{\setbox0=\hbox{$\scriptscriptstyle\rm C$}\hbox{\hbox
to0pt{\kern0.4\wd0\vrule height0.9\ht0\hss}\box0}}}}

\def\bbbz{{\mathchoice {\hbox{$\sf\textstyle Z\kern-0.4em Z$}}
{\hbox{$\sf\textstyle Z\kern-0.4em Z$}} {\hbox{$\sf\scriptstyle
Z\kern-0.3em Z$}} {\hbox{$\sf\scriptscriptstyle Z\kern-0.2em Z$}}}}

\begin{document}

\title{Structure of quantum correlations in momentum space and off diagonal long range order in eta pairing and BCS states}

\author{Paolo Giorda$^{1}$ and Alberto Anfossi$^2$ }

\address{$^1$Institute for Scientific Interchange (ISI), Villa Gualino, Viale Settimio Severo 65, I-10133 Torino, Italy}

\address{$^2$ Dipartimento di Fisica del Politecnico, Corso Duca degli Abruzzi 24, I-10129 Torino, Italy}

\begin{abstract}
The quantum states built with the \emph{eta paring mechanism} i.e., eta pairing states, were first introduced in the context of high temperature superconductivity where they were recognized as important example of states allowing for off-diagonal long-range order (ODLRO). In this paper we describe the structure of the correlations present in these states when considered in their momentum representation and we explore the relations between the quantum bipartite/multipartite correlations exhibited in $k$ space and the direct lattice superconducting correlations. In particular, we show how the negativity between paired momentum modes is directly related to the ODLRO. Moreover, we investigate the dependence of the block entanglement on the choice of the modes forming the block and on the ODLRO; consequently we determine the multipartite content of the entanglement through the evaluation of the generalized \emph{Meyer Wallach} measure in the direct and reciprocal lattice.
The determination of the persistency of entanglement shows how the network of correlations depicted exhibits a self-similar structure  which is robust with respect to ``local" measurements. Finally, we recognize how a relation between the momentum-space quantum correlations and the ODLRO can be established even in the case of BCS states.
\end{abstract}


\maketitle

\section{Introduction}

The eta pairing mechanism for electrons was first introduced by Yang \cite{Yang_eta} in the context of high-temperature superconductivity. Generally speaking, the relevance of the states built by means of the eta pairing mechanism, i.e. eta pairing states, comes from the fact that they are exact eigenstates -- in some cases actual ground-states -- of several relevant models in different fields of condensed matter physics.\\
A first example of a model having the eta pairing states as eigenstates is the Hubbard one. It can be proven \cite{Yang_eta} that in this case these eigenstates cannot be the ground-state since one can build states with different symmetry having a lower energy. Nevertheless, the eta pairing mechanism allows for the formation of Cooper pairs located on a site, rather than separated by a finite coherence length, as in the case of the standard Bardeen, Cooper Schrieffer scheme for superconductivity \cite{BCS}. In this context these eigenstates play a prominent role since they display off-diagonal long-range order  (ODLRO)\cite{Yang_ODLRO}. The latter is a peculiar kind of nonlocal correlation that survive in the thermodynamical limit. It has been shown \cite{Nieh-Su-Zhao} that ODLRO implies Meissner effect and flux quantization, which are both distinctive features of superconducting systems.

On the other hand, for specific regimes of parameters, the eta pairing states turn out to be the ground-state of different extensions of the Hubbard model \cite{DBKS, EKS, MC, AKS, AAS}. In particular, in the bond-charge extension \cite{AAS} a state characterized by the eta pairing mechanism belongs to the lowest-energy sector even at finite positive values of the Coulomb interaction. Also, by mapping the eta pairing state into the spin-$\frac{1}{2}$ language \cite{AKS}, one can recognize it as the ground state of the isotropic Heisenberg ferromagnet, as well as of the Heisenberg antiferromagnet when the anisotropy parameter $\Delta=-1$ \cite{salerno2}.

Due to their relevance, the eta pairing states are natural candidates for the applications of the tools and the paradigms developed within the framework of quantum information theory (QIT) \cite{Nielsen-Chuang,AmicoRMP}. In this context, a central concept is that of entanglement i.e., the quantum correlations among or within subsystems constituting a system. A significant amount of results in QIT aims at the classification and quantification of the existing correlations, quantum or classical, in a given quantum state. The Von Neumann entropy, for instance, measures the quantum correlations of a subsystem with the rest of the system. The quantum mutual information is a measure of the total (quantum and classical) correlations between two subsystems, while the negativity quantifies just the quantum correlations between the latter. One can thus gain a deep insight about the structure of the existing network of correlations in a complex system by evaluating and comparing appropriate entanglement (correlation) measures once the sets of different partitions of the system into subsystems are given.

In the direct lattice picture, Zanardi and Wang \cite{Zanardi-Wang} were the first to apply the above scheme to the eta pairing states by analyzing the entanglement between two sites and showing that the latter vanishes in the TDL. The entanglement between a block of sites and the rest of the lattice was analyzed in \cite{Fan-Lloyd,salerno1,salerno2} where it was shown that it scales logarithmically with the size of the block and how it is connected with the ODLRO.
A through analysis of the entanglement in the direct lattice picture  for pure states and mixtures of eta pairing states was also carried on in \cite{vedraleta}, where it was pointed out that while the two-site entanglement vanishes in the thermodynamical limit (TDL), the two-site classical correlations are still present; when measured by the quantum mutual information, the latter were recognized in \cite{AGM} to coincide with the ODLRO. Finally, in \cite{AGMT,AGM} it was investigated how the quantum (total) correlations of ground-states characterized by the presence of the eta paring mechanism behave at quantum phase transitions. The analysis allows one to classify the latter in terms of two-site or multipartite quantum correlations.

In this paper we aim at studying the entanglement properties of the eta pairing states in their momentum representation. This study on one hand provides a complementary and richer picture of the underlying correlations structure; on the other hand it allows one to explore the relations between the quantum bipartite and multipartite correlations exhibited in $k$ space and the superconducting correlations typical of the direct lattice picture. In particular, we evaluate appropriate measures of correlations considering different choices of the elementary subsystems (single, paired and unpaired modes) and we find how the negativity between paired modes is directly related to the ODLRO present in these states.
Furthermore, the exact diagonalization of the reduced density matrix associated with an arbitrary set of momenta will allow us to study the scaling of the block entropy with the size of the block and to see how it depends on the selection of the modes forming the block and on the ODLRO.  These results will also make it possible to to compare the multipartite entanglement content in the direct and reciprocal lattice picture through the evaluation of the \emph{Meyer Wallach} measure \cite{MeyWalQ,BrennenQ,ScottQ}. The determination of the persistency of entanglement \cite{Briegel} will show how the network of correlations depicted exhibits a self-similar structure  which is robust with respect to ``local" measurements, i.e. measurements of single or paired momentum modes \cite{Greiner}.

In order to compare the results obtained in the eta paring case we finally study the BCS states. Some entanglement properties of these states were studied in \cite{Ent_BCS_Shi, Ent_BCS_Delgado,Ent_BCS_Dunnings,ohkim supercond}. Here, by resorting to the Green's function language developed in \cite{ohkim supercond} we show that while these states exhibit a simpler structure of correlations in momentum space this structure can be directly linked to their ODLRO.

The paper is organized as follows: we start by introducing the $k$-space representation of the eta pairing state then, in section \ref{sec:single-pair-modes}  we analyze the correlations between single modes and pairs of modes. Section \ref{Sec:block-entro} is devoted to the study of the block entropy, while in Section \ref{Sec: Meyer-Wallach} we review the Meyer Wallach measure and we use it to investigate the multipartite content of correlations in the eta pairing states. We conclude the analysis of the correlation properties of the latter in Section \ref{Sec: Persistency} where we analyze the persistency of the entanglement. Finally, in Section \ref{Sec: BCS states} we investigate the connection between entanglement in momentum space and ODLRO in BCS states.

\section{The eta pairing state}\label{Sec: Eta pairing states}

\subsection{Preliminaries}

The eta paring states are built through the action of the so-called eta operator that, in direct lattice picture, is written as
 \be
 \eta^\dagger =\sum_{l=1}^L c_{l\uparrow}^\dagger c_{l\downarrow}^\dagger
 = \sum_{l=1}^L\eta_l^\dagger\;.
 \ee
Here $\eta_l^\dagger=c_{l\uparrow}^\dagger c_{l\downarrow}^\dagger$ is written in terms of $c_{{l} \sigma}^\dagger$ and $c_{{l} \bar{\sigma}}^{\dagger} $ i.e., the fermionic creation operators on the site $l$ of a one-dimensional chain of length $L$; $\sigma \in\{ \uparrow, \downarrow \}$ is the spin label and $\bar{\sigma}$ denotes its opposite.
When acting on the vacuum of the lattice $\bigotimes_{l=1}^L\ket{0}_l$, it creates a pair of particles fully delocalized over a chain of sites of  length $L$. By using the Fourier transform $a^\dagger_{k_j\sigma}\doteq\frac{1}{\sqrt{L}}\sum_{l=1}^{L} e^{i\frac{2\pi}{L}jl}c_{l\sigma}^\dagger$ of each $c_{l\sigma}^\dagger$ one obtains the $k$-space representation of the eta operator:

\be
 \eta^\dagger  =\sum_{j=0}^{L-1} a_{-k_j\uparrow}^\dagger a_{k_j\downarrow}^\dagger= \sum_{j=0}^{L-1}\hat{\eta}_{k_j}^\dagger\;,
\ee
where $\hat{\eta}_{k_j}^\dagger=a_{-k_j\uparrow}^\dagger a_{k_j\downarrow}^\dagger$. Each $k_j=2\pi j/L$ now labels one of $L$ momentum modes, whose local basis is $\mathcal{B}_{k_j}=\{\ket{0}_{k_j}, \ket{\uparrow}_{k_j}, \ket{\downarrow}_{k_j}, \ket{\uparrow\downarrow}_{k_j}\}$. When acting on the
vacuum $\ket{\mbox{vac}}_K=\bigotimes_{j=0}^{L-1}\ket{0}_{k_j}$ the eta operator creates a pair of particles fully delocalized over the whole momentum space. The way in which the delocalization is performed in the two representations is fundamentally different. In the direct lattice each of the $\eta_l^\dagger$ operators acts creating a pair of particles ($\uparrow, \downarrow $) localized on the site $l$. In the $k$ representation, the delocalization is performed already at the level of each $\hat{\eta}_{k_j}^\dagger$; the latter acts creating pairs of particles of the type ($\sigma_{-k_j},\bar{\sigma}_{k_j}$), thus involving the pair of modes $(-k_j,k_j)$.

The generic eta pairing state is obtained by the creation of $N_d$ pairs of
particles:
\be
    \ket{\Psi(L,N_d)}=\sqrt{\frac{(L-N_d)!}{L!N_d!}}(\eta^\dagger)^{N_d} \ket{\mbox{vac}}\;.
\ee

We note that the states considered here are a particular case of the general family of eta pairing states that can be built by operators of the form $\eta^\dagger_\phi=\sum_{l=1}^L e^{i\phi l} c_{l\uparrow}^\dagger c_{l\downarrow}^\dagger$. In the $k$-space picture these operators create pairs with momentum equal to $\phi$. One can see that the structure of correlations of the states generated by the action of $\eta_\phi^\dagger$ does not depend on the particular choice of the value of $\phi$; in the following we will thus choose $\phi=0$.
\\Our analysis is based on the evaluation of the (quantum) correlations between different subsystems i.e., set of momentum modes. To this end we consider the following measures of correlations. The Von Neumann entropy will be used to measure the quantum correlations between a subset $A$ of momentum modes and the rest of the system when the latter is in a pure state.
Its definition is based on the reduced density matrix $\rho_A$:
\be
\mathcal{S}(\rho_A)=\mbox{ tr }(\rho_A \log_2 \rho_A)=-\sum_i \lambda_i \log_2 \lambda_i
\ee
where $\{\lambda_i\}$ are the eigenvalues of $\rho_A$.
In order to measure the \textit{total} i.e, quantum and classical, correlations between two generic subsystems $A$ and $B$ (sets of momentum modes), we use the quantum mutual information \cite{Nielsen-Chuang,vedralrmp,GPW}. This measure is defined in terms of the system's and subsystems' density matrices $\rho_{AB}$, $\rho_{A}$ and $\rho_{B}$:
 \be
\mathcal{I}_{AB}=\mathcal{S}(\rho_{A})+\mathcal{S}(\rho_{B})-\mathcal{S}(\rho_{AB}).\label{QMI}
\ee
The quantum correlations between two generic subsystems $A$ and $B$
will be quantified by the negativity \cite{vidalvernerneg}
\be
 \mathcal{N}(\rho_{AB})=(\|\rho_{AB}^{T_A}\|_1-1)/2;\label{neg}
\ee
where $\rho_{AB}^{T_A}$ is the partial transposition with respect to the subsystem $A$ applied on $\rho_{AB}$, and $\|O\|_1\doteq Tr\sqrt{ O^\dagger O}$ is the trace norm
of the operator $O$.
All the above functionals properly capture the bipartite correlations between two subsystems. For the multipartite correlations we will make use of the \emph{Meyer Wallach} measure \cite{MeyWalQ,BrennenQ,ScottQ}, see section \ref{Sec: Meyer-Wallach} for the definition.

As mentioned in the introduction, the eta pairing states were first recognized as relevant since they allow for superconducting correlations i.e., ODLRO, which for these states are defined as:
\be\label{ODLRO}
    \lim_{r\rightarrow\infty}\langle \eta^\dagger_j  \eta_{j+r}\rangle = n_d(1-n_d)\, .
\ee
where $j, j+r$ label two sites of the lattice at distance $r$.
A first bridge between ODLRO and a measures of correlations defined in the context of quantum information theory was established in \cite{AGM} where the analysis carried on in the direct lattice framework allowed to recognize that these correlations are simply proportional to  $\mathcal{I}_{j,j+r}$ i.e., the mutual information between two generic sites $j,j+r$, where $r$ is an arbitrary distance .

\subsection{Correlations in $k$-space: single and pairs of modes.}
\label{sec:single-pair-modes}

In this section we begin to analyze the structure of correlations of the eta pairing states at the local level. The calculations involved in the study of both the single mode and the pairs of modes will lead in the following sections to the generalization to the case of block of modes.

\subsubsection{Single mode}

We start the analysis by evaluating the reduced density matrix $\rho_{k_j}$ of the single generic mode $k_j$.
The calculations are instructive since some of the features that hold for the single mode case will be useful for the determination of the reduced density matrix of the set of modes we consider in the following sections.\\
It turns out that the only non-vanishing matrix elements of $\rho_{k_j}$ are
the diagonal ones. The latter are the expectation values of the following projectors:
\begin{eqnarray*}
    a_{k_j\downarrow} a_{k_j\uparrow} a^\dagger_{k_j\uparrow} a^\dagger_{k_j\downarrow} &=& |0\rangle_{k_jk_j}\langle 0|, \\
    a^\dagger_{k_j\uparrow} a_{k_j\uparrow} a_{k_j\downarrow} a^\dagger_{k_j\downarrow} &=& |\uparrow\rangle_{k_jk_j}\langle \uparrow|,\\
    a^\dagger_{k_j\downarrow} a_{k_j\downarrow} a_{k_j\uparrow} a^\dagger_{k_j\uparrow} &=& |\downarrow\rangle_{k_jk_j}\langle \downarrow| ,\\
    a_{k_j\uparrow}^\dagger a^\dagger_{k_j\downarrow} a_{k_j\downarrow} a_{k_j\uparrow} &=& |\uparrow\downarrow\rangle_{k_jk_j}\langle \downarrow\uparrow|\;.
\end{eqnarray*}
Indeed, the state $\ket{\Psi(L,N_d)}$ is built through the action of superpositions of operators $\hat{\eta}_{k_j}^\dagger$: every time the mode $k_j$ is occupied, the $-k_j$ one is changed correspondingly. Therefore, since the off-diagonal elements correspond to projectors that represent processes that, acting locally onto the $k_j$ mode, leave unchanged the $-k_j$ one they have zero expectation value.
Accordingly, the diagonal single-mode reduced density matrix reads:
\be
    \rho_{k_j} = \mbox{diag}\{AA',AB,AB,BB'\}
\ee
where
\bae
    \left.\begin{array}{l}
        A=\frac{L-N_d}{L} \\[5pt]
        A'=\frac{L-N_d-1}{L-1}
    \end{array}\right\}
    \begin{array}{c}
        L\rightarrow \infty \\
        \longrightarrow \\
        \\
    \end{array} &a& = 1-n_d\;.\nonumber\\
    \left.\begin{array}{l}
        B=\frac{N_d}{L}\\[5pt]
        B'=\frac{N_d-1}{L-1}
    \end{array}\right\}
    \begin{array}{c}
        L \rightarrow \infty\\
        \longrightarrow\\
        \\
    \end{array} &b& = 1-a=n_d\;.
\eae
Capital letters refer to finite-size expressions while lower cases refer to their asymptotic expression for $L \rightarrow \infty$. We can now evaluate as first measure of correlations the Von Neumann entropy of $\rho_{k_j}$:
\be
    \mathcal{S}_{k_j}=-2\left(a\log a +b\log b\right)
\ee
which gives the amount of total (quantum) correlations that the single generic mode has with the rest of the system. Just as much as in the direct lattice picture (\cite{AGMT,AGM}), the correlations are directly determined by the filling $n_d=n/2$. and they reach their maximal value in correspondence with the {\em half filling} case $n=1$.

\subsubsection{Two modes}

We now consider the correlations of the subsystem  constituted by two generic modes $(k_i,k_j)$. We have to distinguish between two cases. For $k_i\neq -k_j$ the reduced density-matrix $\rho_{_{-k_i,k_j}}$ is diagonal with respect to the local basis $\mathcal{B}_{k_i}\otimes\mathcal{B}_{k_j}$; indeed, as described for the single mode case, the off-diagonal elements correspond to expectation values of projectors that change the state of the modes $k_i$, $k_j$ and do not affect the modes $-k_i$, $-k_j$; they hence have zero expectation value. For large $L$, the eigenvalues are $a^{4-\alpha} b^\alpha$, $\alpha=0,\dots, 4$, each one appearing with multiplicity given by $m_\alpha={4 \choose m}$.
This scheme can be generalized to a higher number modes, as we shall see in section \ref{Sec:block-entro}.
The case $k_i=-k_j$ has to be treated separately. The support of the reduced density-matrix is the subspace spanned by
\be
    \mathfrak{B}_{-k_j,k_j}= \{\ket{0,0}_j,\, \ket{\uparrow,\downarrow}_j,\, \ket{\downarrow,\uparrow}_j, \,
    \ket{\uparrow\downarrow,\uparrow\downarrow}_j\}
    \label{base(-kk)}\;,
\ee
where $\ket{\alpha,\beta}_j\doteq\ket{\alpha}_{-k_j}\ket{\beta}_{k_j}$. Indeed, the sole states that can be built by the action of the $\hat{\eta}^\dagger_{k_j}$ operators belong to this subspace. Thus the reduced density matrix of the subsystem $(-k_j,k_j)$ has just a $4\times 4$ nonzero sub-block relative to the subspace spanned by (\ref{base(-kk)}):
\be
    \rho_{|_{\mathfrak{B}_{(-k_j,k_j)}}}=\left(
    \begin{array}{cccc}
     a^2 & 0  & 0  & 0   \\
     0   & ab & ab & 0   \\
     0   & ab & ab & 0   \\
     0   & 0  & 0  & b^2 \\
    \end{array}\right)\;,\label{Eq: twopairedmodes rho}
\ee
whose diagonal form is $\mbox{diag}\{a^2,2ab,0,b^2\}$.

The density-matrices just determined allow us to calculate $i)$ the quantum correlations between the pair of modes with the rest of the system $ii)$ the total and quantum correlations between  any pair of modes $(k_i,k_j)$. We have that:
\bae
 \mathcal{S}_{k_i,k_j} &=& \left\{\begin{array}{lr}
  -2\left(a\log{a}+b\log{b}+ab \right) & k_i=-k_j\\
  -4\left(a\log{a}+b\log{b}\right) & k_i\neq -k_j
  \end{array}\right.\label{VNEkikj}\\
 \mathcal{I}_{k_i,k_j} &=& \left\{\begin{array}{ll}
  \mathcal{S}_{k_i}+2ab & k_i=-k_j\\
  0 & k_i\neq -k_j
 \end{array}\right.\label{MIkikj}\\
 \mathcal{N}_{k_i,k_j} &=& \left\{\begin{array}{ll}
  ab/3 & k_i=-k_j\\
  0 & k_i\neq -k_j
  \end{array}\right.\label{Nkikj}\;,
\eae
where: $\mathcal{S}_{k_i,k_j}$,  the Von Neumann entropy of $\rho_{_{k_i,k_j}}$, measures the quantum correlations between the pair of modes and the rest of the systems. $\mathcal{I}_{k_i,k_j}$, the mutual information, and $\mathcal{N}_{k_i,k_j}$, the negativity, measure the total and quantum correlations between the modes.

\subsubsection{Paired modes}

In order to understand the $k$-space network of correlations, at the local (and multipartite) level, we turn to the natural extension of the above calculations by considering the four modes $(-k_j,k_j), (-k_i,k_i)$. This will allow us, on one hand, to identify as elementary subsystem the generic pair of modes $(-k_i,k_i)$ and to study the nature of the correlations between such kind of subsystems, i.e. \emph{two-pair} correlations. On the other hand, the strategy at the basis of the calculations can be generalized to an arbitrary number of pairs of modes, thus enabling the evaluation of the entropy of a block of modes with respect to the rest of the system (block entropy).

We first must determine the reduced density-matrix $\rho_{_{(-k_i,k_i, -k_j,k_j)}}$. It turns out that the latter has support only on the subspace spanned by $\mathfrak{B}_{_{-k_i,k_i,-k_j,k_j}}=\mathfrak{B}_{-k_i,k_i}\otimes \mathfrak{B}_{-k_j,k_j}$, see (\ref{base(-kk)}). In fact, the only $4$-modes projectors $\mathbb{P}_\alpha^4$ that have a non vanishing expectation value are those that preserve the number $\alpha$ of $(\sigma_{-k_j},\bar{\sigma}_{k_j})$ pairs of particles (as in the two-mode case). Moreover, the expectation values are all equal for a given $\alpha$ and can be straightforwardly determined as functions of $\alpha$. For large $L$ we have:
\bae
    \langle \mathbb{P}_0^4 \rangle &=& a^4\quad \langle \mathbb{P}_1^4 \rangle = a^3b \nonumber \\
    \langle \mathbb{P}_2^4 \rangle &=& a^2b^2\quad \langle \mathbb{P}_3^4 \rangle = ab^3\quad \langle \mathbb{P}_4^4 \rangle = b^4\;, \label{4mod-prj}
\eae
i.e., $\langle \mathbb{P}_\alpha^4 \rangle=a^{4-\alpha}b^\alpha$ is the expectation value of the pair-preserving projector between local states containing $\alpha$ pairs. The fact that the expectation values of the elements involving a given number of pairs are all equal allows for a simple expression for the of diagonal form of the non vanishing sub-block  of $\rho_{_{(-k_i,k_i, -k_j,k_j)}}$:
\be
    \rho_{|_{ \mathfrak{B}_{(-k_i,k_i,-k_j,k_j)}}} = \mbox{diag}\{m_0 a^4,m_1a^3b,m_2a^2b^2, m_3ab^3,m_4 b^4 \}\;.
\ee
The coefficients $m_\alpha={4 \choose \alpha}$ are given by the dimension of the square sub-matrix of $\rho_{(-k_i,k_i,-k_j,k_j)}$ that corresponds to a fixed number of pairs. The latter is, in fact, equal to the number of ways one can set $\alpha$ pairs in $4$ modes.

We have now all the ingredients to derive the following correlations measures:
\begin{eqnarray}
    \mathcal{S}_{(-k_i,k_i,-k_j,k_j)} &=& -4\left\{a\log{a}+b\log{b}+1/2a(1-a) \cdot \right.\nonumber \\
    &&\cdot \left.\left[4+ a(1-a)(3 \log_2{3} - 5)\right]\right\} \label{VNEkkhh}\\
    \mathcal{I}_{(-k_i,k_i),(-k_j,k_j)} &=& 2a(1-a)\left[2+ a(1-a)(3 \log_2{3} - 5)\right]\nonumber \\\label{MIkkhh}\\
    \mathcal{N}_{(-k_i,k_i),(-k_j,k_j)} &=& 0\label{Nkkhh}
\end{eqnarray}
Furthermore, the above evaluated reduced density matrices allow us to determine that the generic single mode $k_j$ is uncorrelated with respect to any pair of modes $(-k_i,k_i)$, i.e. $\mathcal{I}_{k_j,(-k_i,k_i)}=0$.
\begin{figure}[!h]
    \begin{centering}
    \fbox{\includegraphics[height=6cm, width=4cm, viewport= 20 0 670
    510, clip]{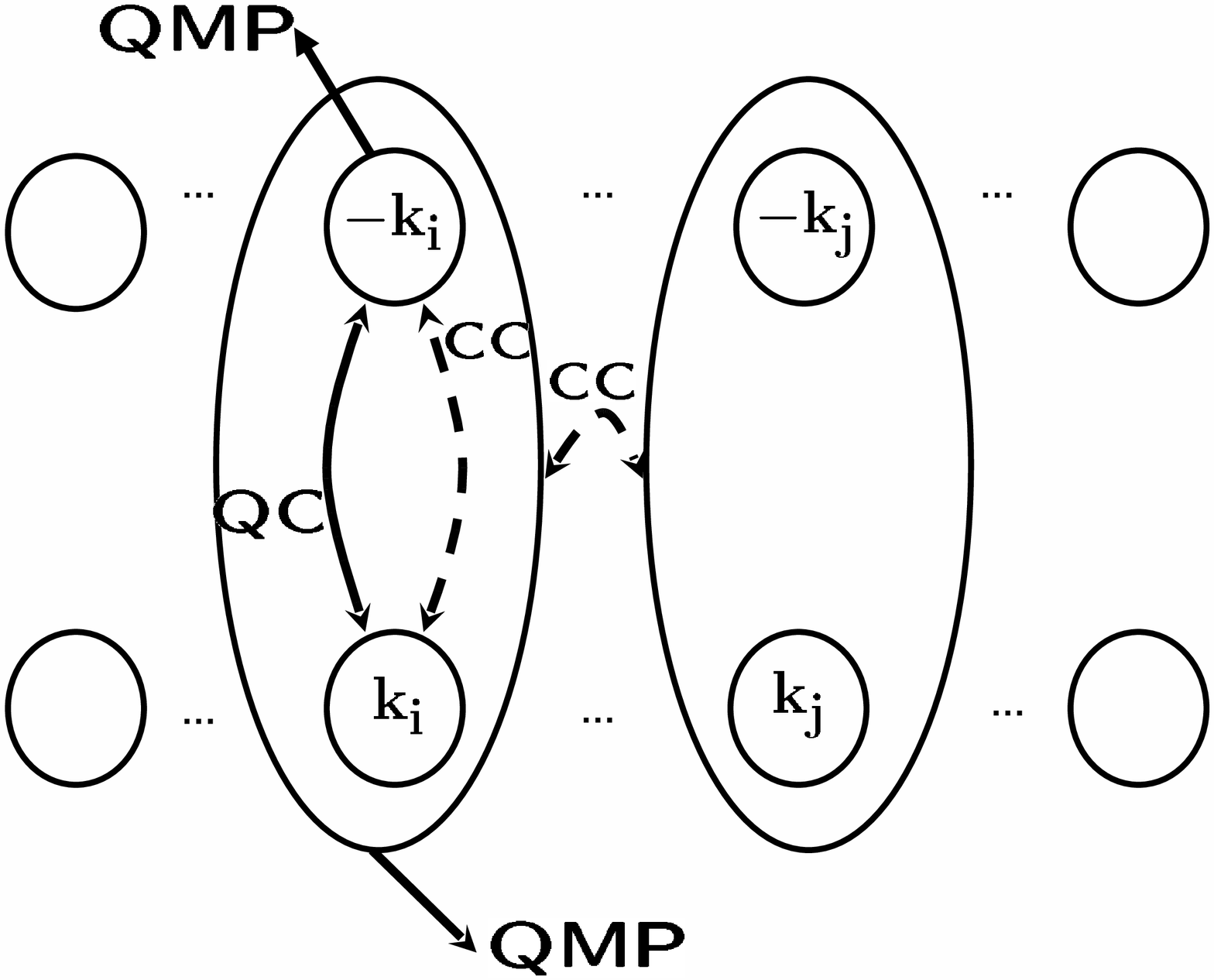}}
    \fbox{\includegraphics[height=6cm, width=4cm, viewport= 0 0 680
    500, clip]{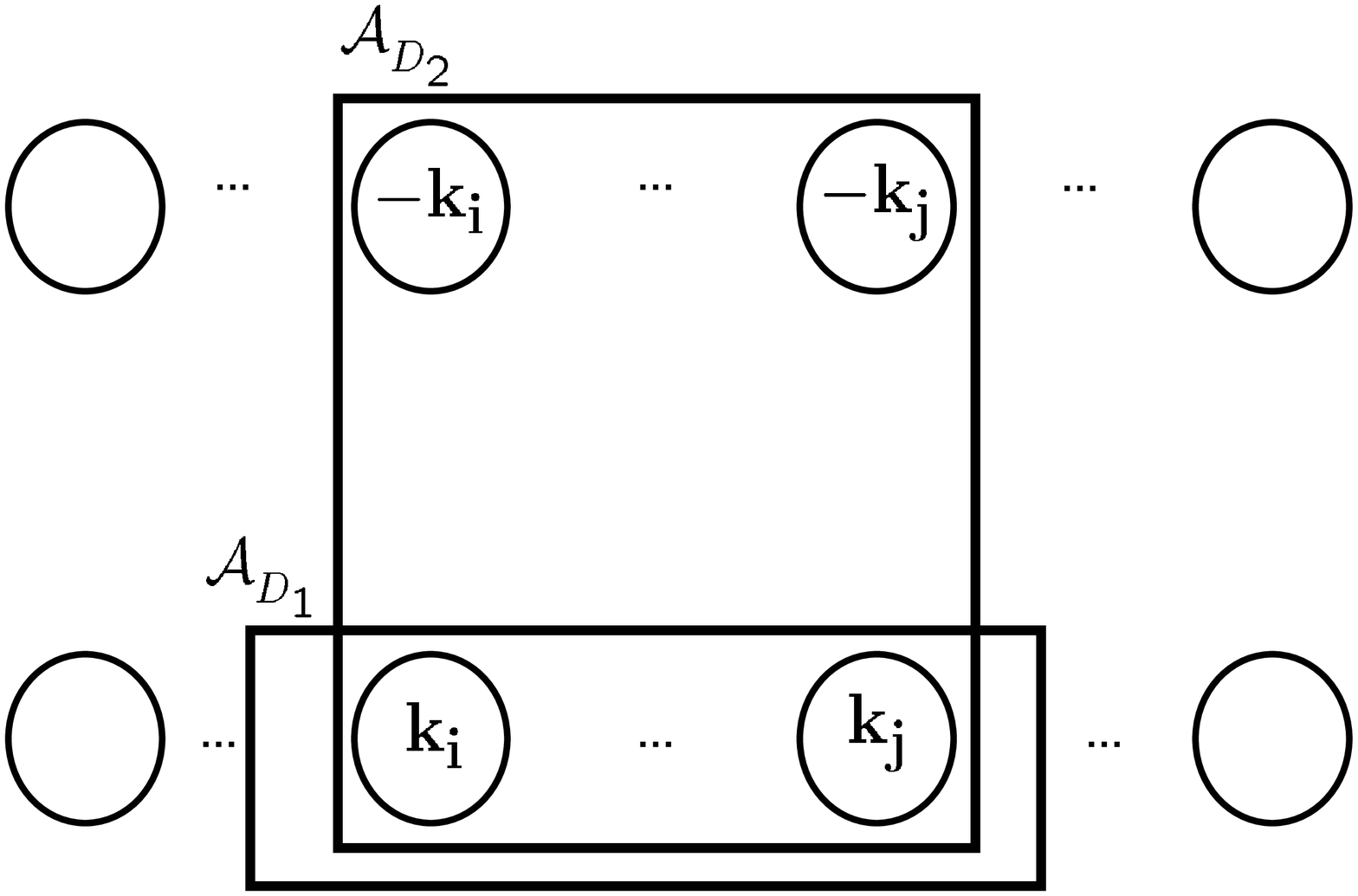}}
    \caption{Left: schematic representation of the correlations of the generic single mode $-k_i$ and of the paired modes $(-k_i,k_i)$. QC stands for quantum correlations, CC for classical correlations, while QMP stands for quantum multipartite. Right: two opposite ways of grouping the modes in blocks i.e.,  paired ($\mathcal{A}_{D_1}$) and unpaired ($\mathcal{A}_{D_2}$) modes.}
    \label{Fig:corr}
    \end{centering}
\end{figure}

\subsubsection{Discussion}\label{sec:discKspace}

The above $k$-space calculations suggest the following scenario sketched in figure \ref{Fig:corr}. The quantum correlation of the single mode $k_j$ with the rest of the system have both a two-mode and a multipartite nature. The mode has two-point quantum correlations only with the mode $(-k_j)$, see (\ref{Nkikj}), while all the other two-mode (total) correlations are zero: when $k_i\neq -k_j$, $\mathcal{I}_{k_j,k_i}=0$.

The fact that $\mathcal{S}_{-k_j,k_j}\neq 0$, see (\ref{VNEkikj}), implies the existence of a quantum multipartite contribute to the single-mode correlations. The nature of such a contribute is better characterized by the fact that the single mode is uncorrelated with any pair of modes $k_i,k_l\neq -k_j$, i.e., $\mathcal{I}_{k_j,k_ik_l}=0$. This means that we are in presence of $n$-way entanglement, with $n>3$. All the considered measures reach their maximum at $a=1/2$; the latter condition is satisfied when $n_d=n/2=1/2$. We now focus our attention on the elementary subsystems constituted by the pairs of modes $(-k_j,k_j)$. These are quantum correlated with the rest of the system (\ref{VNEkikj}) and in order to describe the multipartite correlations related to them we can introduce the notion of \emph{multi-pair} entanglement; indeed, the mutual information (\ref{MIkkhh}) indicates the existence of \emph{two-pair} correlations which are of classical nature only, since at the same time, the \emph{two-pair} quantum correlations, as measured by the negativity (\ref{Nkkhh}), are zero. This kind of distribution of multipartite correlations in terms of two-pair classical ones is the same encountered in the direct lattice picture where a the level of two-sites the correlations are just classical and proportional to the ODLRO.

It is now important to explore the nature of some of the described correlations and highlight their relations to meaningful physical quantities. In particular it turns out that it is possible to establish a direct connection of the quantum correlations between two paired modes and the ODLRO. Indeed, the negativity $\mathcal{N}_{-k_j,k_j}$ is calculated on the basis of the off -diagonal elements of the non-vanishing part of the reduced density matrix of the paired modes (\ref{Eq: twopairedmodes rho}). These elements coincide with  expectation values $\average{\hat{\eta}_{k_j}^\dagger \hat{\eta}_{-k_j}}=\average{\hat{\eta}_{-k_j}^\dagger \hat{\eta}_{k_j}}$ and by writing them in terms of direct lattice operators one has:
\be
\mathcal{N}_{k_j,-k_j}=\average{\hat{\eta}_{k_j}^\dagger \hat{\eta}_{-k_j}}/3=
\frac{1}{3 L^2}\sum_{l\neq m}\average{\eta_l^\dagger \eta_m}+O(\frac{1}{L})
\label{Eq: Neg vs ODLRO eta}
\ee
The sum has $L(L-1)$ non vanishing terms and we see that {\em the quantum correlations between paired momentum modes are given by the average of the pairing correlations in the direct lattice}. Since the latter do not depend on the lattice sites' indexes $l,m$ and are all equal to (\ref{ODLRO}) we the above equation establishes a direct connection between {\em the quantum correlations between paired momentum modes and the sufficient condition for superconductivity i.e., ODLRO}.
As a final remark we note that the ODLRO (\ref{ODLRO}) explicitly appears in the expression of other correlation measures, in particular those indicating, as previously discussed, the presence of multi-pair entanglement (\ref{VNEkikj}) and its reflection at the level of two-pairs correlations(\ref{MIkkhh}).
In the following sections we will see how this multi-pair nature of quantum correlations determines the behaviour of the block entanglement and of the related measure of multipartite entanglement and how these can be again expressed in terms of ODLRO.

\subsection{Block entropy}
\label{Sec:block-entro}

We now deepen our study of the structure of the correlations in the eta-paring
state $\ket{\Psi(N_d)}$ by considering how the subsystems constituted by blocks of modes are entangled with the rest of the system: \emph{block entropy}.\\
In the direct lattice picture the eta-pairing state can be mapped, through a particle-hole transformation \cite{AKS}, onto the ground state of the isotropic ferromagnetic Heisenberg model. In this framework, by taking advantage of the permutational invariance of such states, one can evaluate the entropy of blocks of \emph{sites} \cite{salerno1,salerno2,BrandaoEntOrderParam}.\\
Here, in the $k$-space representation, the structure of correlations between blocks of \emph{modes} is more complex and allows for a richer picture: as we shall see in the next section, the general form of the block entropy depends on the particular choice of the set of modes constituting the block.

\subsubsection{Unpaired modes}

In this section we evaluate the block entropy between a set $\mathcal{A}_{D_1}$ of $D_1$ single modes and the rest of the system such that $\forall k_j\in\mathcal{A}_{D_1} \;\Rightarrow\; -k_j \notin \mathcal{A}_{D_1}$. For this choice of set of modes the reduced density matrix $\rho_{D_1}$ of the block is diagonal since the off-diagonal elements correspond to projectors that act only on the $k_j$-mode while leaving the $-k_j$ mode unchanged and therefore have zero expectation value. The generic diagonal element is the expectation value of the product of $D_1$ local projectors $\mathbb{P}_{k_j}\in\{\ketbra{0}{0},\ketbra{\uparrow}{\uparrow}, \ketbra{\downarrow}{\downarrow}, \ketbra{\uparrow\downarrow}{\uparrow\downarrow}\}$. The result can be written in terms of the number of modes and the total number of fermions $M$ involved by the projectors:
\be
    \langle \mathbb{P}^{D_1}_M\rangle = \frac{{L-2D_1 \choose N_d-M}}{{L \choose N_d}} \underRarr{\small L \rightarrow \infty} a^{2D_1-M}b^M\;.
\ee
There are ${2D_1\choose M}$ projectors that involve $M$ particles and they all have the same above expectation value. Thus the von Neumann entropy of the block corresponding to the set $\mathcal{A}_{D_1}$ can be written as
\bae
    \mathcal{S}_{D_1} &=& -\sum_{M=0}^{2D_1}{2D_1\choose M}a^{2D_1-M}b^M \log(a^{2D_1-M}b^M)\nonumber\\
    &=& -2D_1 \left(a\log a + b\log b \right) = D_1\mathcal{S}_{k_j} \label{sing_blockentropy}\;.
\eae
Indeed, this result reflects the factorization of the density matrix of the block
\be
    \rho_{D_1}=\otimes_{k_j\in\mathcal{A}_{D_1}}\rho_{k_j}\label{fact-rho-kj}\;,
\ee
thus extending the result (\ref{VNEkikj}) for the single mode $k_j$. Equation (\ref{fact-rho-kj}) shows how the modes belonging to $\mathcal{A}_{D_1}$ are completely uncorrelated among each other. The correlations of the block are given by $i)$ multipartite correlations between each mode and the rest of the system $ii)$ the two-mode correlations: the boundary of the block cuts the links between each $k_j\in\mathcal{A}_{D_1}$ and its complementary $-k_j\notin\mathcal{A}_{D_1}$, see figure \ref{Fig:corr}.

\subsubsection{Paired modes}

In this section we consider a different situation which is opposite, in some sense, with respect to the previous one. We evaluate the block entropy between a set $\mathcal{A}_{D_2}$ of $D_2$ single modes and the rest of the system such that $\forall k_j\in\mathcal{A}_{D_2} \;\Rightarrow\; -k_j \in \mathcal{A}_{D_2}$.

The results obtained in section \ref{sec:single-pair-modes} for one and two pairs of modes can be extended to an arbitrary number $D_2/2$ of pairs of modes $(-k_j,k_j)$. In particular, the expressions (\ref{4mod-prj}) obtained for the two pairs of modes can be easily generalized. The projectors $\mathbb{P}_\alpha^{D_2}$ that have nonvanishing expectation value are those that conserve the number of pairs of particles $(\sigma_{-k_j},\bar{\sigma}_{k_j})$ and that represent, as explained above, processes that ``coherently" affect the $(-k_j,k_j)$ pair.\\
The density matrix relative to $D_2/2$ pairs of modes is block-diagonal. Each block can be labeled by the fixed number of pairs $\alpha \in [0,D_2]$ of $(\sigma_{-k_j},\bar{\sigma}_{k_j})$ particles involved by $\mathbb{P}_\alpha^{D_2}$. The elements of a given block are all equal and their explicit expression is
\be
    \big\langle \mathbb{P}_\alpha^{D_2}\big\rangle = \frac{{L-D_2\choose N_d-\alpha}} {{L\choose N_d}}
    \begin{array}{c}
        L \rightarrow \infty\\
        \longrightarrow\\
        \\
    \end{array}
    a^{D_2-\alpha}b^\alpha\;.
\ee
The dimension of the block is $m_\alpha = {D_2 \choose \alpha}$, i.e., it corresponds to the number of ways one can place $\alpha$ pairs of particles $(\sigma_{-k_j},\bar{\sigma}_{k_j})$ in $D_2/2$ pairs of modes. Each block labeled by $\alpha$ has $m_\alpha-1$ null eigenvalues. In the large-$L$ limit the nonvanishing part of the spectrum of the reduced density matrix is then given by the following set of $D_2$ eigenvalues:
\be
    \{m_0 a^{D_2},m_1 a^{D_2-1}b, \dots, m_{D_2-1}ab^{D_2-1},m_{D_2} b^{D_2}\}\;.
    \label{Eq: Eigval Binomial}
\ee
Since $b=1-a$ we see that the eigenvalues follow a binomial distribution and the Von Neumann entropy thus reads
\begin{eqnarray}
    \mathcal{S}_{D_2} &=& -\sum_{M=0}^{D_2}{D_2 \choose M}a^{D_2-M}b^M\log\left[ {D_2 \choose M}a^{D_2-M}b^M\right]\nonumber
    \\ \label{pairs_blockentropy}\,.
\end{eqnarray}

For large values of $D_2$  the latter expression was evaluated in \cite{knessl} and it has the following asymptotic expression:
\begin{equation}
    \mathcal{S}_{D_2} \sim \frac{1}{2}\log(D_2)+\frac{1}{2}\log(2\pi ab) \;,\label{TDL_pairs_blockentropy}
\end{equation}
that holds for $ab\gg 0$.

\subsubsection{Paired and un paired modes}

We have seen that the choice of the set of modes crucially determines the (asymptotic) behavior of the block entropy. In particular,
the block entropy of the set of modes $\mathcal{A}_{D_1}$, which does not contain any pair $(-k_j,k_j)$, grows linearly with the number of modes (\ref{sing_blockentropy}). On the contrary, in the opposite situation, the block entropy of the set of modes $\mathcal{A}_{D_2}$ grows only logarithmically.\\
We note that results (\ref{pairs_blockentropy}) and (\ref{TDL_pairs_blockentropy}) are only \emph{formally} equivalent to the one obtained in the spin-models context \cite{salerno1} since the latter refer to the direct lattice picture and they take into account completely different subsystems  (blocks of sites). Indeed, while in the direct lattice the choice of the block of modes is unambiguous here, in the $k$-space, the block entropy behaves logarithmically only for a specific choice of the set of modes, i.e, $\mathcal{A}_{D_2}$. Furthermore, in order to give a complete description of the block entropy one has to consider the intermediate picture of a block composed by $D=D_1+D_2$ modes belonging to the set
$\mathcal{A}_{D_1+D_2}=\mathcal{A}_{D_1}\bigcup\mathcal{A}_{D_2}$, where $D_1$ is the number of unpaired modes $k_j\in\mathcal{A}_{D_1}$ and $D_2/2$ is the number of the paired modes $(-k_j,k_j)\in\mathcal{A}_{D_2}$. The generic nonzero element of the reduced density matrix of the block is given by the expectation value of projectors of the type
\be
    \mathbb{P}_D(D_1,F,\alpha)=\mathbb{P}_{D_1}(F-2\alpha)\otimes \mathbb{P}_{D_2}(\alpha)\;.
\ee
The latter are characterized by a fixed number of fermions $F=M+2\alpha$:
$M$ is the number of single particles involved by the $D_1$ single mode projectors $\mathbb{P}_{k_j}\in\{\ketbra{0}{0},\ketbra{\uparrow}{\uparrow}, \ketbra{\downarrow}{\downarrow}, \ketbra{\uparrow\downarrow}{\uparrow\downarrow}\}$, while $\alpha$ is the number of pairs of particles $(\sigma_{-k_j},\bar{\sigma}_{k_j})$ involved by the $D_2$ projectors relative to the pairs of modes.\\
It turns out that the generic expectation value can be written as the product of expectation values of $\mathbb{P}_{D_1}(F-2\alpha)$ and $\mathbb{P}_{D_2}(\alpha)$, i.e.
\be
    \begin{array}{clc}
        \langle\mathbb{P}_D(D_1,F,\alpha)\rangle & = & \langle\mathbb{P}_{D_1}(F-2\alpha)\rangle \cdot\langle\mathbb{P}_{D-D_1}(\alpha)\rangle= \\
        & & \\
        =\frac{{L-D_1-D\choose N_d-F+\alpha}}{{L\choose N_d}}
        & = & \frac{{L-2D_1\choose N_d-F+2\alpha}} {{L\choose N_d}} \frac{{L-D+D_1\choose N_d-\alpha}}{{L\choose N_d}}\;.\\\label{eigv-tensor}
    \end{array}
\ee
This expression can be used to recognize that, for large $L$  the contributions of paired and unpaired modes can be factorized; indeed, for finite $L$, the reduced density matrix of the block can then be written as:
\be
    \rho_{D}=\rho_{D_1} \otimes \rho_{D_2}
\ee
and this implies that for any set $\mathcal{A}_{D_1}$ the unpaired modes are uncorrelated with respect to any finite number $D_2$ of paired modes $(-k_j, k_j)$. If we now let $L$ take arbitrarily large values formulas (\ref{eigv-tensor}) can be simply expressed in terms of $a,b=1-a$ as
\bae
    \langle\mathbb{P}_D(D_1,F,\alpha)\rangle &=& a^{D+D_1-F+\alpha}b^{F-\alpha}=\nonumber \\
    &=& \left(a^{2D_1-F+2\alpha} b^{F-2\alpha}\right) \left(a^{D-D_1-\alpha}b^\alpha\right) \nonumber \\
\eae
and one can use this relation to recognize again that [see (\ref{MIkikj})] for $L$ large enough the unpaired modes are uncorrelated among each other, i.e.
\be
    \rho_{D_1}=\bigotimes_{k_j\in\mathcal{A}_{D_1}} \rho_{k_j}\;.\label{rho-tensorprod}
\ee
Since for fixed $D_1$ and $D_2$ the structure of $\rho_{D_1}$ and $\rho_{D_2}$ is the same analyzed in the previous sections the evaluation of the eigenvalues $\rho_{D}$ follows along the same lines and the Von Neumann entropy of the block of modes in the set $\mathcal{A}_{D_1+D_2}$ can thus be simply expressed as the sum of the block entropies pertaining to $\mathcal{A}_{D_1}$ and $\mathcal{A}_{D_2}$:
\bae
    \mathcal{S}_{D_1+D_2}&=& -2D_1\left(a\log a+b\log b\right) \nonumber\\
    &-&\sum_{M=0}^{D_2}{D_2 \choose M}a^{D_2-M}b^M \log\left[ {D_2 \choose M}a^{D_2-M}b^M\right]\;.\nonumber\\
\eae
When $D_2$ is sufficiently large on can again approximate the binomial distribution and obtain:
\be
    \mathcal{S}_{D_1+D_2} \sim -2D_1\left(a\log a + b \log b \right) +\frac{1}{2}\log{D_2}\;.\label{block-entro-mix}
\ee

\subsubsection{Discussion}

The above results suggest the following interpretation in terms of two-point and multipartite entanglement. On one hand the linear growth of the block entropy $\mathcal{S}_{D_1}$ has two main contributions: $i)$ the two-mode quantum correlations of each mode $k_j\in\mathcal{A}_{D_1}$ with the corresponding mode $-k_j\notin\mathcal{A}_{D_1}$;
$ii)$ the multipartite correlations that a single mode shares with the rest of the system.
On the other hand the logarithmic growth $\mathcal{S}_{D_2}$ cannot be described in terms of two-pair quantum correlations: since $\mathcal{N}_{(-k_j,k_j),(-k_i,k_i)} = 0$, the growth of $\mathcal{S}_{D_2}$ reflects the multipartite (multi-pair) nature of the correlations of the subsystem $\mathcal{A}_{D_2}$ with the rest of the system. In the intermediate case $\mathcal{A}_{D_1}\cup\mathcal{A}_{D_2}$, these two effects sum and  $\mathcal{S}_{D_1+D_2}$ accounts for both the two-pair and the multi-pair contributes to correlations.
We finally observe that the binomial distribution (\ref{Eq: Eigval Binomial}), that corresponds to the eigenvalues of the density matrix relative to block of pair of modes, contains the link with the ODLRO. Indeed, the variance of the distribution is just $D_2 n_d(1-n_d)$, thus to a higher amount of ODRLO it corresponds a broader probability distribution and consequently a higher value of the block entropy (\ref{pairs_blockentropy}). The above discussed multi-pair contribution to the entanglement present in the momentum picture is thus directly related to the superconducting correlations. An analogue relation was find in the direct lattice picture in \cite{salerno1,salerno2,BrandaoEntOrderParam,Fan-Lloyd} where the entropy of block of sites was considered and an analogous binomial distribution for the eigenvalues of the reduced density matrix was found. In the following we will compare the two pictures on the basis of the Q-measure of entanglement.

\subsection{Q-measure of entanglement}\label{Sec: Meyer-Wallach}

We use the above results to compute the Meyer-Wallach measure of multipartite entanglement $Q(\ket{\psi})$. The latter was first introduced in \cite{MeyWalQ} for multi-qubit pure states. In \cite{BrennenQ} it was shown how $Q$ can be simply expressed as the average  linear entropy of the single qubits:
\beq
    Q(\ket{\psi})=2\left(1-\frac{1}{n}\sum_{k=1}^{n} tr \rho^2_k \right)\;,
\eeq
here $\rho_k$ is the reduced density matrix of the $k$-th of $n$ qubits. This expression of $Q$ allowed in \cite{ScottQ} for two different kinds of generalization. On one hand, one can extend the measure to the case of multi-qudit states, i.e. to quantum systems that are composed by $n$ identical \emph{elementary} subsystems $\mathcal{S}$. Each of the latter lives in $\mathcal{H}_d$ ($\mbox{dim}(\mathcal{H}_d)=d$), while the whole system lives in $(\CC^d)^{\otimes^n}$. In this case, $Q$ measures the average linear entropy of the subsystems $\mathcal{S}$. On the other hand, one can further generalize and extend the measure by considering \emph{principal} subsystems $\mathcal{A}_m$ composed by $m$ elementary subsystems $\mathcal{S}$. Here the average must be extended to all the $C^n_m = n!/m!(n-m)!$ possible inequivalent choices of $\mathcal{A}_m$ and it can be written as
\beq
    Q_{m,d} = \frac{d^m}{d^m-1}\left(1- \frac{1}{C^n_m}\sum_{\vec{i}} \mbox{tr} \rho^2_{\vec{i}}  \right)\; ,
\label{Qmeas}
\eeq
where each $\vec{i}= \{i_1,\dots,i_m\}$ is a subset of indexes identifying the $m$ elementary subsystems $\mathcal{S}$ composing a given $\mathcal{A}_m$, the latter being characterized by its reduced density matrix $\rho_{\vec{i}}$. The factor $d^m/(d^m-1)$ normalizes the measure to one.

$Q_{m,d}$ is thus a multipartite entanglement measure based on the average of a bipartite one (linear entropy): it quantifies the average entanglement between blocks of qudits and the rest of the system. Anyway, only as the size $m$ of the blocks increases, $Q_{m,d}$ becomes really sensitive to correlations of an increasing global nature \cite{ScottQ,deOliveira}. This property is in general difficult to exploit since it requires the evaluation of the reduced density matrices of blocks of arbitrary dimension. As we have seen in the previous sections this turns out to be possible in $k$-space.

\subsubsection{Q-measure in $k$-space}

\begin{figure}[!h]
    \begin{centering}
    \fbox{\includegraphics[height=5cm, width=6cm, viewport= 0 0 710
    480, clip]{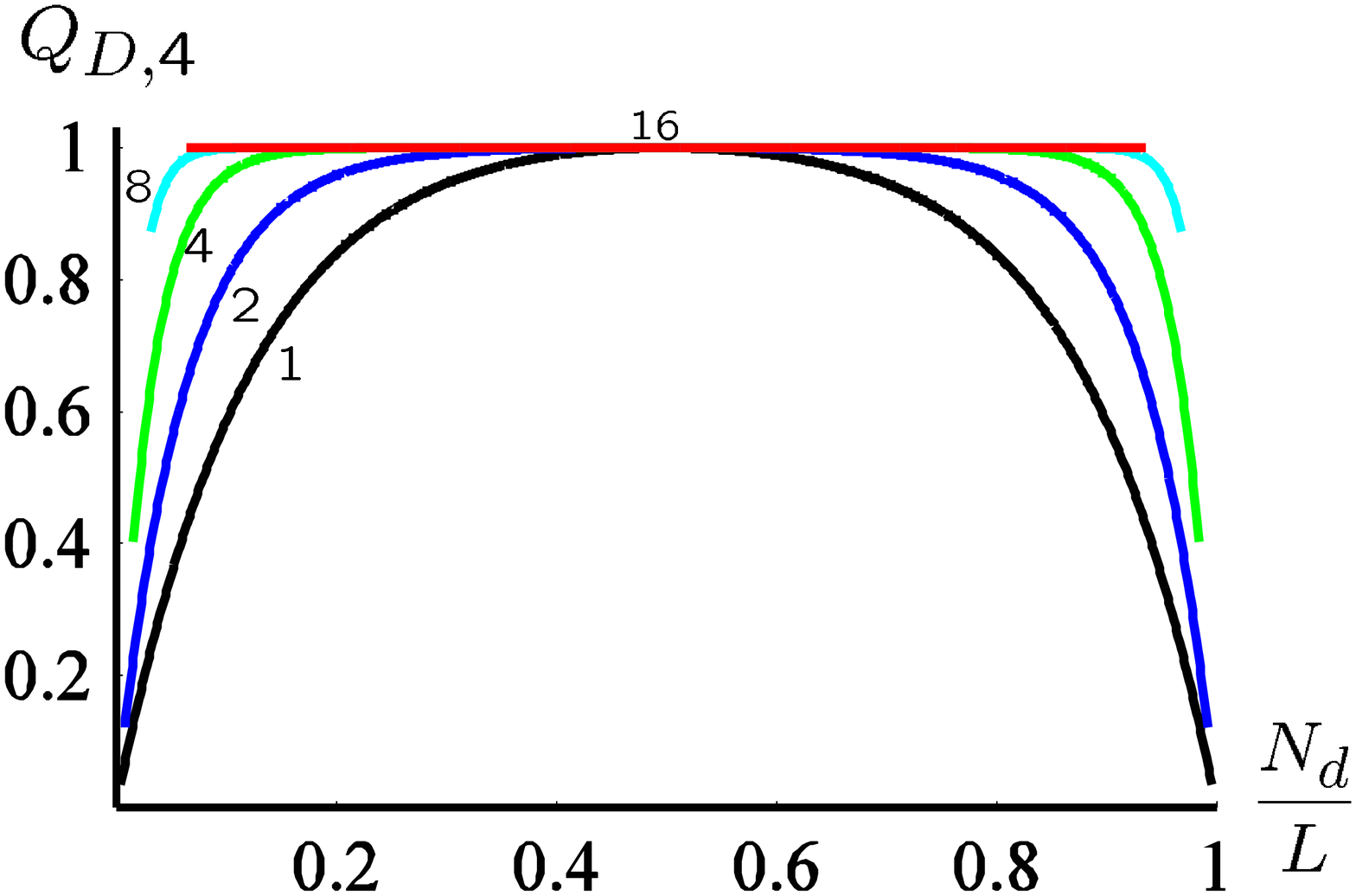}}
    \caption{(Color online) Plot of the measure of multipartite entanglement
    $Q_{D,4}$ in $k$-space for different sizes of the elementary subsystem; the size varies form bottom to top: $D=1,2,4,8,16$\, ($L=1000$)}
    \label{Fig:KspaceMWRegIII}
    \end{centering}
\end{figure}

In this section we analyze the multipartite correlations in $k$-space as they
can be described by $Q_{m,d}$.
We have to specialize (\ref{Qmeas}) to the $k$-space case. $d=4$ is the dimension of the single-mode Hilbert space. Each principal subsystem $\mathcal{A}_{D=D_1+D_2}$ is composed by an even number $D$ of modes;
$D_1$ are unpaired modes $k_i$ while $D_2/2$ are pairs of modes $(-k_j,k_j)$. For a given $L$, one has ${L\choose D}$ inequivalent choices. In order to evaluate the density matrix $\rho_D$ of the generic $\mathcal{S}_D$ one has to take into account its composition in terms of paired and unpaired modes.
As we have seen in section \ref{Sec:block-entro}, in the finite size case, $\rho_D$ can be written as a tensor product $\rho_{D_2}\otimes\rho_{D_1}$ but the unpaired modes are not uncorrelated, i.e. $\rho_{D_1}\neq\bigotimes_{i}\rho_{k_i}$. For fixed $D_2$ the various $\rho_D$ have the same spectrum and since Tr$(\rho_D^2)= \mbox{Tr}(\rho_{D_2}^2) \mbox{Tr}(\rho_{D_1}^2)$, $Q$ can eventually be written as
\begin{equation}
    Q_{D,4} = \frac{4^D}{4^D-1}\left[1-{L\choose D}^{-1} \sum_{D_2=0}^D f(D_2)\mbox{Tr}(\rho_{D_1}^2)\mbox{Tr}(\rho_{D_2}^2) \right]\;,\label{KspaceMW}
\end{equation}
where
\be
    f(D_2)=\frac{\prod_{i=0}^{D_1-1}(L-2i)}{D_1!} \frac{\prod_{j=0}^{D_2/2-1}(L-2D_1-2j)}{D_2!}\label{MW-kspace}
\ee
is the number of equivalent partitions of $D$ into $D_1=D-D_2$ single modes and $D_2/2$ pairs of modes;
\bae
    \mbox{Tr}(\rho_{D_1}^2) &=& \sum_{M=0}^{2D_1}{2 D_1\choose M} \left[\frac{{L-2D_1\choose N_d-M}}{{L \choose N_d}}\right]^2,\\
    \mbox{Tr}(\rho_{D_2}^2) &=& \sum_{\alpha=0}^{D_2}\left[{D_2\choose \alpha} \frac{{L-D_2\choose N_d-\alpha}}{{L \choose N_d}}\right]^2.\label{Tr-MW-kspace}
\eae
Formula (\ref{KspaceMW}) can be numerically evaluated for $L$ large; in figure \ref{Fig:KspaceMWRegIII} we plot $\; Q_{D,4}$ for different values of $D$ as a function of $N_d/L$. As $D$ grows, the measure, which is normalized to one,  rapidly saturates to its maximal value for any filling. This result confirms the analysis of the structure of correlations carried on in the previous sections. Indeed, in the $k$-space picture, \emph{multipartite entanglement is the dominant feature of the eta pairing state}.\\

\subsubsection{Q-measure in direct lattice}

We conclude by considering the extension of the above analysis to the direct lattice picture. As already mentioned, through a particle-hole transformation it is possible to translate the eta-pairing states into the spin language. At variance with the $k$-space picture, where the idea of considering the pairs of modes $(-k_j,k_j)$ is natural and necessary, in the direct lattice the choice of the set of elementary subsystems (the sites of the chain) constituting the generic principal subsystem (the block of sites) is unambiguous. The local Hilbert space has effective dimension $d=2$, since each site can be either empty or doubly occupied. Due to permutational symmetry, each block of sites has the same $\rho_D$ and, consequently, the average linear entropy coincides with the linear entropy of one block.
\begin{figure}[!h]
    \begin{centering}
    \fbox{\includegraphics[height=5cm, width=6cm, viewport= 0 0 710
    480, clip]{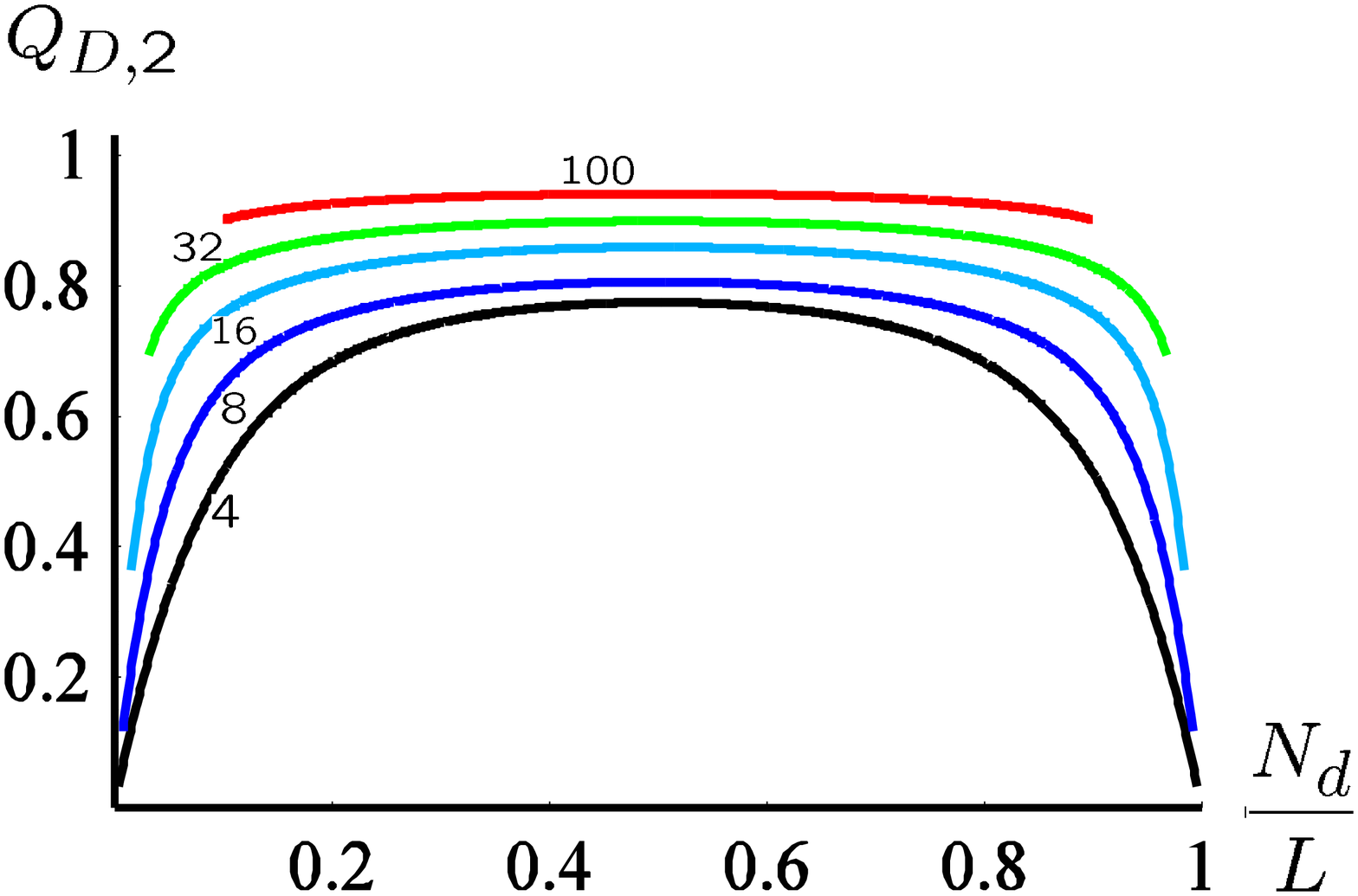}}
    \caption{ (Color online) Plot of the measure of multipartite entanglement
    $Q_{D,2}$ in direct lattice picture for different sizes of the elementary subsystem; the size varies form bottom to top: $D=4,8,16,32,100$\, ($L=1000$)}
    \label{Fig:DirMWRegIII}
    \end{centering}
\end{figure}
In figure \ref{Fig:DirMWRegIII} it is shown how, as $D$ grows,  $Q_{D,2}$ asymptotically approaches its  maximal value. This result confirms that in the direct lattice picture the multipartite entanglement does play a fundamental role in the eta-pairing states \cite{vedraleta,AGM}.

It is now interesting to compare the structure of correlations scenario emerging from the two frameworks: direct and reciprocal lattice.\\
In the direct lattice there are no two-point quantum correlations: the two-site concurrence is zero when $L \rightarrow \infty$ for any pair of sites, thus implying a vanishing entanglement ratio \cite{AGM}. This is a signature that the nature of the correlations in this picture is basically multipartite. Furthermore, the two-point correlations, i.e. the mutual information between two sites, since they do not depend on the distance between the sites are uniformly spread over all the chain; as previously mentioned their value is proportional to the ODLRO.

In the $k$-space picture the only ``two-point" quantum correlations --as measured by negativity-- are those shared by each mode $k_j$ with its ``complementary" $-k_j$ and we have seen how they are directly connected with ODLRO. Furthermore, the total two-point correlations are localized between these two complementary modes ($\mathcal{I}_{k_i,k_j}\neq 0\iff k_i=-k_j$). As discussed in section \ref{sec:discKspace}, it is possible even from this coarse-grained level of description, infer the relevance of the multipartite kind of entanglement.\\
We can thus pass to analyze the main differences at the level of multipartite correlations, as measured by $Q$. The comparison of figure \ref{Fig:KspaceMWRegIII} and figure \ref{Fig:DirMWRegIII} points out that in general, for a given size $D$ of the principal subsystem (i.e. block of sites/modes), the value of $Q$ is higher in the $k$-space case. Indeed, in the direct lattice for fixed $D$, $Q_{D,2}$ is proportional to the linear entropy
\be
    1-\sum_{l=0}^D\left[{D\choose l}{L-D\choose N_d-l}/{L\choose N_d}\right]^2\;.\label{dirLinEnt}
\ee
In the $k$-space, $Q_{D,4}$ is the average of terms that take into account the composition of the blocks in paired and unpaired modes and it is of course greater than the least of these terms. The latter corresponds to the block composed by $D/2$ paired modes and formally coincides with (\ref{dirLinEnt}), see (\ref{MW-kspace}-\ref{Tr-MW-kspace}). These considerations suggest that while the dominant feature of the eta pairing state is the multipartite entanglement, this kind of correlations play a major role in k-space representation. The difference between the two representations manifests itself in the distinct asymptotic behavior of the block entropy in the two pictures: in direct space $\mathcal{S}_D$ can grow at most logarithmically with the size of the block, while in $k$-space it can grow linearly (\ref{block-entro-mix}).

\subsection{Persistency of entanglement}\label{Sec: Persistency}

We conclude our analysis of the structure of correlations in the eta pairing states by briefly discussing --both in direct and in reciprocal lattice-- the operational effort required to destroy all the entanglement present in the system. To this end compute the \emph{persistency of entanglement}  $\mathcal{P}_e$ that was introduced in \cite{Briegel} to test the strength of quantum correlations present in a state in view of its use for quantum information protocols. For a $L$-qudit pure state $\mathcal{P}_e$ is defined as the minimum number of local measurements that reduces the entangled state to a product state of the $L$ qudits (i.e. that disentangles the state).

In the direct lattice we have seen that the local state is a qubit, since the $j$-th site can only be either empty or doubly occupied. Due to the structure of the eta pairing states, the effect of a single local measure $\mathbb{P}_\alpha=\otimes_{i\neq j}\openone_i\otimes|\alpha\rangle_{jj}\langle\alpha|$, $\alpha\in\{0,\uparrow\downarrow\}$ is to reduce the number of sites over which the eta pairs can be delocalized from $L$ to $L-1$ and to generate a new eta pairing state with either $N_d-1$ or $N_d$ pairs, i.e.,
\bae
    \mathbb{P}_0\ket{\Psi(L,N_d)}/p_0&=&
    |0 \rangle_{jj}\langle 0|\otimes\ket{\Psi(L-1,N_d)}\label{meas_0} \\
    \mathbb{P}_{\uparrow\downarrow}\ket{\Psi(L,N_d)}/p_{\uparrow\downarrow}&=&
    |\uparrow\downarrow \rangle_{jj}\langle \uparrow\downarrow| \otimes\ket{\Psi(L-1,N_d-1)}\label{meas_2} \nonumber \\
\eae
where $p_\alpha=\mbox{Tr}(\mathbb{P}_\alpha\ketbra{\Psi(L,N_d)}{\Psi(L,N_d)})$.
The eta pairing states thus display a self-similar behavior under local measurements. Moreover, they turn out to be robust to noise since for $N_d\le L/2$ [$N_d\ge L/2$] one needs to perform at least $N_d$ [$L-N_d$] measures in order to destroy all the quantum correlations present in the state, i.e.,
\be
    \mathcal{P}_e[\ket{\Psi(L,N_d)}]=\left\{\begin{array}{ll}
            N_d & N_d\le L/2\\
            L-N_d & N_d\ge L/2\;.
            \end{array}\right.
\ee
Indeed, it easily seen that the minimum number of measures needed to factorize $\ket{\Psi(N_d)}$ is reached by repeating $N_d$ measures of type (\ref{meas_0}), if $N_d\le L/2$, or $L-N_d$ measures of type (\ref{meas_2}), if $N_d\ge L/2$.

Despite the fact that in momentum space, being the local space four-fold, the number of possible local measures is four $\mathbb{P}_\alpha=\otimes_{i\neq j}\openone_i\otimes|\alpha\rangle_{jj}\langle\alpha|$, $\alpha\in\{0,\uparrow,\downarrow,\uparrow\downarrow\}$, the situation is similar to the direct lattice one, since the more convenient sequence of measures to factorize the state consists again in performing just measures of type $\mathbb{P}_{\uparrow\downarrow}$ when $N_d\le L/2$ or of type $\mathbb{P}_0$ when $N_d\le L/2$. In $k$-space, however, the effect of one of these local measures is to reduce the available space for the delocalization of the eta pairs from $L$ to $L-2$ and to generate an eta pairing state, consisting of either $N_d$ or $N_d-2$ pairs. Accordingly, here the persistency of entanglement $\mathcal{P}_e^{(k)}$ reads
\be
    \mathcal{P}_e^{(k)}[\ket{\Psi}(L,N_d)]=\left\{\begin{array}{ll}
            N_d/2 & N_d\le L/2\\
            (L-N_d)/2 & N_d\ge L/2\;.
            \end{array}\right.\;.
\ee
In the perspective of the robustness of the entanglement present in the eta pairing states, the direct lattice representation seems therefore to be the more favorable, since $\mathcal{P}_e=2\mathcal{P}_e^{(k)}$. More interestingly, the eta pairing states, both in $k$-space and in the direct lattice, display a self-similar behaviour under local measures. In particular, once a local measure is performed, the output state is characterized by the same structure of correlation of the input state, both at the level of bipartite and multipartite entanglement, as at the level of the ODLRO present in the state.

We finally note that recently the possibility of measuring correlations between paired momentum modes in a similar context have been experimentally achieved, as reported in \cite{Greiner}.

\section{BCS states}\label{Sec: BCS states}

The direct relation between the entanglement in the momentum picture and the ODLRO that has been highlighted for the case of the eta pair states suggests to explore whether analogue relations hold for other states exhibiting superconducting correlations.
In this last section we thus pass to analyze the structure of quantum entanglement in momentum space for the first example of state exhibiting superconducting correlations: the BCS state \cite{BCS}. We will in particular see how the desired direct relation can be established.\\
The BCS state was introduced as an Ans\"{a}tz wave function defined in momentum representation for models which allow the formation of Cooper pairs:
\bae
\ket{\mbox{BCS}}&=&\prod_{\bf k} (u_{\bf k}+v_{\bf k} a_{{\bf k}\uparrow}^\dagger a_{-{\bf k}\downarrow}^\dagger)\ket{\mbox{vac}}=\otimes_{\bf k} \ket{\Psi_{\bf k}}\nonumber\\
\eae
Here the electron pair is created localized in the pair of modes ${\bf k},-{\bf k}$ by the fermionic operators $a_{{\bf k}\uparrow}^\dagger a_{-{\bf k}\downarrow}^\dagger$. The overall state is a grand-canonical one since the number of particle is not fixed and it is normalized: for each $k$ the coefficients satisfy $u_{\bf k}^2+v_{\bf k}^2=1$.
One can easily see that, at variance with the eta-pair states, the structure of correlations is quite simple since no multipartite entanglement is present: the state is factorized in momentum space and the only (quantum) correlations exhibited are those between the paired mode ${\bf k},-{\bf k}$. Indeed, the concurrence between the modes ${\bf k}$ and $-{\bf k}$ can be evaluated by treating the $\ket{\Psi_{\bf k}}$ state as effectively qubit states and it coincides with the negativity: $ \mathcal{C}_{\bf k,-k}= 2u_{\bf k} v_{\bf k}=\mathcal{N}_{\bf k,-k}$ \cite{Dunning_vonDelft}.
We now describe how the above entanglement properties can be related to the fundamental property of BCS states i.e., ODLRO. In order to accomplish this task we can simply resort to some of the calculations explicitly derived in \cite{ohkim supercond}. There the two particle density matrix $\rho^{(2)}$ has been obtained using the language of Green functions \cite{ManyBodyQFT} and it has been used  to study the spin entanglement properties of two electrons forming a Cooper pair. The explicit expression of $\rho^{(2)}$ also allows for the the determination of the ODLRO present in the system. We now briefly recall, for the sake of clarity, the main ingredients needed for the discussion and we finally identify the relevant relations that allow one to link the quantum properties to the basic superconducting correlations.

The two-electron  space-spin density matrix can be written as:
\bae
\rho^{(2)}(x_1,x_2,x_1',x_2')&\doteq&\frac{1}{2}
\average{\hat{\psi}^\dagger(x_1')\hat{\psi}^\dagger(x_2')
\hat{\psi}(x_1)\hat{\psi}(x_2)}\nonumber\\
&=&-\frac{1}{2}G(x_1t_1,x_2t_2,x_1't_1^+,x_2't_2^+)\nonumber\\
\eae
Here we have that $x_i=({\bf x_i},s_i)$ and $\hat{\psi}^\dagger(x_i) (\hat{\psi}(x_i))$ is the creation (destruction) operator for a particle in position ${\bf x_i}$ with spin $s_i=\pm 1/2$; the expectation value is taken on the ground state at zero temperature.
Moreover, the Green function $G(x_1t_1,x_2t_2,x_1't_1^+,x_2't_2^+)=G_{(1,2;1',2')}$ is defined in terms of the creation [annihilation] operator $\hat{\psi}_H^\dagger(x_i) [\hat{\psi}_H(x_i)]$ in the Heisenberg representation as:
\be
G_{(1,2;1',2')}\doteq
\average{T[\hat{\psi}_H^\dagger(x_1t_1)\hat{\psi}_H^\dagger(x_2t_2)
\hat{\psi}_H(x_2't_2^+)\hat{\psi}_H(x_1't_1^+)]}\;,
\ee
where $T$ is the time-ordering operator and $t_i^+$ is intended as a temporal instant following $t_i$ and infinitesimally close to it. Standard  arguments \cite{ManyBodyQFT} show that $G_{(1,2;1',2')}$ can now be can be factorized in terms of single particle Green's functions in such a way to encompass the presence of pairs of electrons in a bound state:
\be
G_{(1,2;1',2')}=G_{(1,1')}G_{(2,2')}-G_{(1,2')}G_{(2,1')} -F_{(1,2)}F_{(1',2')}^\dagger\;.
\ee
Here the important part for our discussion is now the anomalous Green's function:
\bae
F^\dagger_{(1,2)}=-i\average{\hat{\psi}_H^\dagger(x_1t_1)\hat{\psi}_H^\dagger(x_2t_2)}
\eae
that accounts for the pairing mechanism and is composed by the product of a $1/2$-spin part $I_{s,s'}=i\sigma_y$ and a spatial part $F({\bf x}_1 t_1,{\bf x}_2 t_1)$. The latter is of fundamental importance since it is responsible for the ODLRO present in the system; it thus is the ingredient in which one can find the connection between the entanglement properties of the BCS state in the momentum representation and its superconducting correlations. Indeed, the spatial part of $F^\dagger$, just as much as the spatial part of $G_{(i,j)}$, depends only on the differences $\Delta{\bf x}_{1,2}={\bf x}_1-{\bf x}_2$, $\delta_i=t_i-t_i'$ and it can be evaluated when $\delta_i\rightarrow 0$ as:
\bae
i F^\dagger(\Delta{\bf x}_{1,2})&=&\frac{1}{V}\sum_{\bf k} u_k v_k \exp{[{\bf k}({\bf x}_1-{\bf x}_2)] }\nonumber\\
&=& \frac{1}{2 V}\sum_{\bf k} \mathcal{N}_{\bf k,-k} \exp{[{\bf k}({\bf x}_1-{\bf x}_2)] }.\label{Eq: Neg vs ODLRO BCS}
\eae
The last line comes from the observation that the $F^\dagger({\bf x}_1-{\bf x}_2)$ can be written as Fourier series whose ${\bf k}$ components are just given by the two mode concurrence previously introduced $\mathcal{N}_{\bf k,-k}$. We finally recall that \cite{ohkim supercond}, after taking the continuum limit and integrating in the ${\bf k}$ variable, the limit of the two-particles space-spin density matrix when $|{\bf x}_i-{\bf x}_i'|$ going to infinity  defines the ODLRO:
\be
\frac{1}{2}I_{s_1,s_2}I_{s_1',s_2'}F({\bf x}_{1,2})F^*({\bf x}_{1',2'}).
\ee
We thus see that, in analogy to what we have described in the eta-pair case, in BCS state {\em the superconducting correlations that imply Meissner effect and flux quantization are functionally related to the entanglement properties of the state in the momentum representation}.

We can finally resume the scenario emerging from the comparison between the structure of correlations of the two kind of states considered in this paper.
In both cases it possible to establish a direct, tough different, link between quantum correlations in $k$-space and superconducting correlations, see (\ref{Eq: Neg vs ODLRO eta}) and (\ref{Eq: Neg vs ODLRO BCS}). However, while the quantum correlations between paired modes play a prominent role in both kind of states, the relations found for the eta pairing states between the ODLRO and the multipartite correlations suggest that in order to establish a superconducting order in lattice models that have such states as eigenstates, more resources in terms of multipair/multipartite correlations are needed.

\section{Conclusions}

In this paper we addressed the task of describing the structure of the correlations in the $k$-space picture and their relations with the superconducting correlations (ODLRO) for two important example of states: the eta-pairing and the BCS states. These two kind of states share as relevant elementary subsystems the {\em paired} modes $(-k_j,k_j)$ and it turns out that in both cases the quantum correlations between these modes can be directly related with the existing ODLRO. In particular, in the eta-paring case the negativity between two arbitrary paired modes can be written as the average of the pairing correlations that defines ODLRO. As for the BCS state, the Green's function formalism developed in \cite{ohkim supercond} allows to recognize how the ODLRO can be written as a Fourier transform of the negativity between the paired modes.\\
While the momentum description of the BCS states is fairly simple, since the only existing correlations are of bipartite nature, the eta-pairing states require a more articulated analysis. The latter unveiled a network of correlations which is much richer also with respect to the one expressed by the same states in the direct lattice picture. Part of the multipartite content of the entanglement present in these states is shown to have a {\em multi-pair} nature and to be directly determined by the amount of ODLRO present in the system.
The exact evaluation of the reduced density matrix of arbitrary large blocks of modes enabled the study of the entanglement that these subsystems have with the rest of the system (block entanglement).  In this respect, the peculiar structure of correlations in the $k$-space, at variance with the direct lattice picture, implies that the behaviour of the block entanglement in the limit of  large numbers of modes heavily depends on the peculiar choice of the modes that constitute the subsystem. While for blocks composed by $D_1$ {\em unpaired} modes the block entanglement grows linearly with $D_1$, in the case of $D_2$ {\em paired} modes the spectrum of the reduced density matrix is given by a binomial distribution, whose variance is given by the ODLRO, implying the block entanglement to grow only logarithmically with the number of modes $D_2$.
The analysis allows one to assess also the intermediate pictures, where an arbitrary number $D=D_1+D_2$ of modes is considered, and thus to determine the measure of multipartite entanglement firstly devised for qubits by Meyer and Wallach in \cite{MeyWalQ}. This measure, applied both in direct picture and in the $k$-space one, shows that in both cases the multipartite content of the entanglement of the eta pairing states is indeed their dominant feature and it is thus at the basis of their superconducting properties. In view of a possible application to the eta-paring states for quantum computational tasks, we finally show how the evaluation of the persistency of entanglement points out a self-similar structure of the correlations in these states which is robust with respect to ``local" measurements both in the momentum and in direct lattice representation.\\

\begin{acknowledgements}
The authors would like to thank M.Rasetti, A. Montorsi, M. Keyl, D. Schlingemann and
C. Knessl for useful comments and discussions.
\end{acknowledgements}

\end{document}